\documentclass[aps,prl,twocolumn,showpacs,amsmath,amssymb,superscriptaddress]{revtex4}

\usepackage{graphicx}
\usepackage{bm}

\renewcommand{\vec}{\mathbf}
\renewcommand{\v}{\vec}
\renewcommand{\d}{\mathrm{\,d}}

\renewcommand{\theta}{\vartheta}

\begin{document}

\title{Nonlocal Nonlinear Optics in cold Rydberg Gases}

\author{S. Sevin\c{c}li}
\affiliation{Max Planck Institute for the Physics of Complex Systems, 01187 Dresden,Germany}
\author{N. Henkel}
\affiliation{Max Planck Institute for the Physics of Complex Systems, 01187 Dresden,Germany}
\author{C. Ates}
\affiliation{School of Physics and Astronomy, University of Nottingham, Nottingham, NG7 2RD, United Kingdom}
\author{T. Pohl}
\affiliation{Max Planck Institute for the Physics of Complex Systems, 01187 Dresden,Germany}

\begin{abstract}
We present an analytical theory for the nonlinear optical response of a strongly interacting Rydberg gas under conditions of electromagnetically induced transparency. Simple formulae for the third order optical susceptibility are derived and shown to be in excellent agreement with recent experiments. The obtained expressions reveal strong nonlinearities, which in addition are of highly nonlocal character. This property together with enormous strength of the Rydberg-induced nonlinearities is shown to yield a unique laboratory platform for nonlinear wave phenomena, such as collapse-arrested modulational instabilities in a self-defocussing medium. 
\end{abstract}


\maketitle
Advances in designing materials with highly intensity-dependent refraction \cite{pba02,cpa04,rbc06} have ushered in numerous studies of nonlocal nonlinear wave phenomena \cite{wkb02,cpa03,mml06,ssk07,bpa09}. Many of these settings, such as, nematic liquid crystals \cite{pba02,cpa04} or thermal media \cite{rbc06}, require high  power laser light. On the other hand, electromagnetically induced transparency (EIT) in ultracold multi-level atoms \cite{fim05,hhd99} provides an elegant mechanism to suppress photon loss and simultaneously increase light-matter interaction times to enhance nonlinear effects. Combined with sufficiently large nonlinearities, this holds great potential for few-photon nonlinear optics \cite{hfi90,bhb09} and may enable applications in communication and quantum information science.

Recently, it was recognized that EIT-schemes involving highly excited atomic Rydberg levels provide promising perspectives for such applications \cite{frpe05,pode10,olle10,moja07,mami07,moba08,zhzh09,scgu10,prma10,gbe10}. In particular, the huge polarizability of Rydberg states gives rise to giant Kerr coefficients \cite{moba08}, but also entails strong long-range interactions, which render Rydberg-EIT media intrinsically nonlinear. Indeed, a recent theory for two-photon pulses revealed the emergence of strong effective photon-photon interactions \cite{gof11}, while experiments \cite{prma10} and numerical calculations \cite{atse11} demonstrated greatly enhanced nonlinear absorption coefficients in the opposite limit of large photon numbers.

\begin{figure}
	\includegraphics[width=.99\columnwidth]{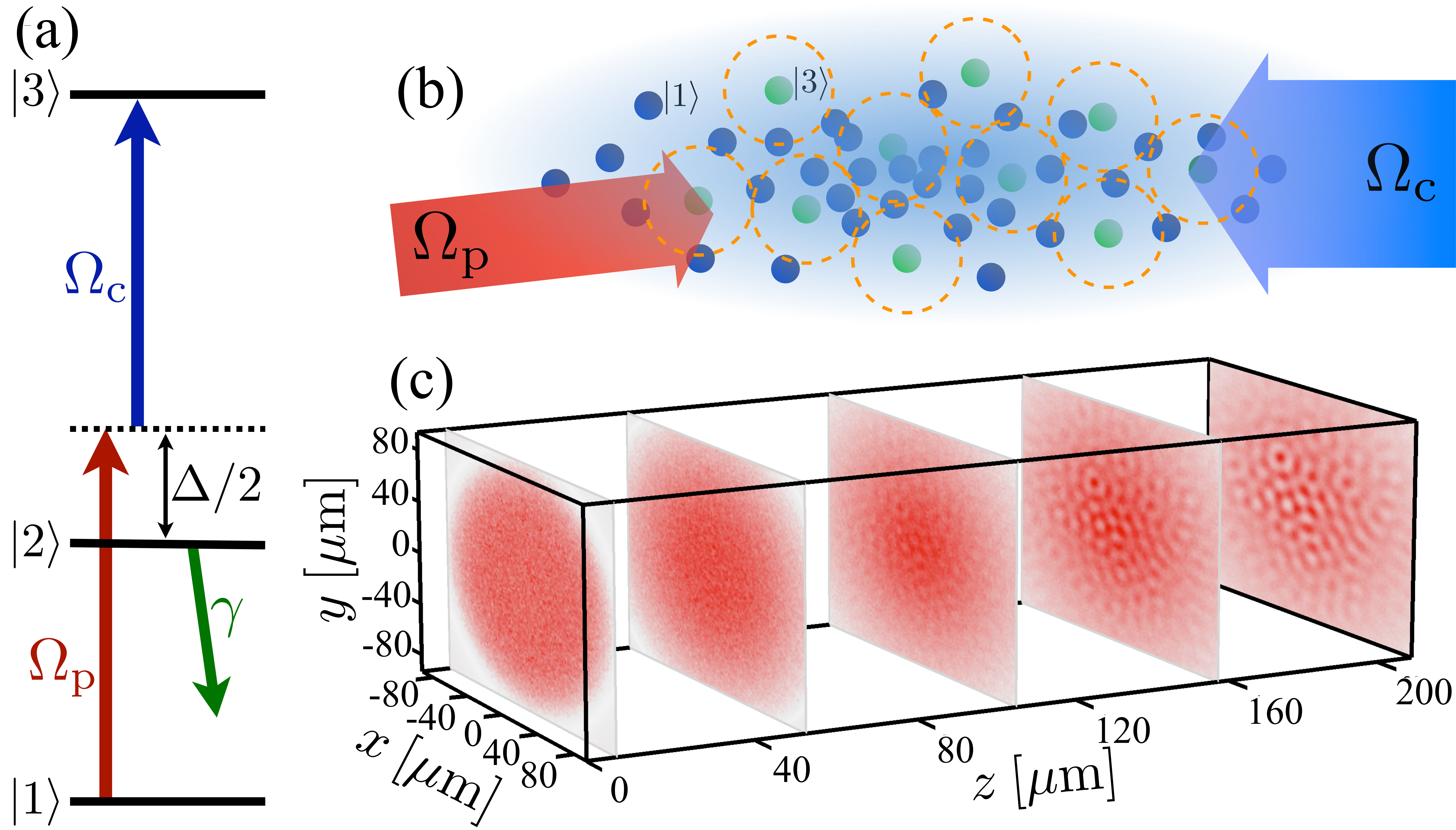}
\caption{a) Three-level scheme for isolated atoms, where the atomic ground state $|1\rangle$, an intermediate state $|2\rangle$ and a highly excited Rydberg state $|3\rangle$ are mutually driven by a strong control and a weak probe field with Rabi frequencies $\Omega_{\rm c}$ and $\Omega_{\rm p}$, respectively. On two-photon resonance, EIT ensures lossless propagation of the optical fields, unaffected by spontaneous decay ($\gamma$) and the single-photon detuning $\Delta/2$. (b) In a gas of atoms, the strong van der Waals interaction between atoms in Rydberg states ($|3\rangle$) inhibit multiple Rydberg excitations within a blockade radius $R_c$, giving rise to a strongly nonlinear optical response of the medium.  The resulting nonlinear beam propagation, for example, leads to modulation instabilities, as shown in (c) for a Rubidium $70S_{1/2}$ Rydberg gas with a density of $8\times10^{13}$cm$^{-3}$ and $\Omega_{\rm p}/2\pi=0.35$MHz, $\Omega_{\rm c}/2\pi=80$MHz, $\Delta/2\pi=1.2$GHz.}
\label{fig1}
\end{figure}

In this letter, we develop an analytical theory for the nonlinear optical response of a strongly interacting Rydberg-EIT medium to monochromatic multi-photon light sources. Based on the approach, we give a simple formula for the nonlinear absorption coefficient that provides an excellent description of recent measurements on cold Rubidium gases \cite{prma10}. For large single-photon detunings, absorption is shown to be greatly suppressed -- yet maintaining huge refractive nonlinearities, that exceed previous records in ultracold Kerr media \cite{hhd99} by several orders of magnitude. Combined with their long range this makes for an ideal nonlinear medium to study nonlocal wave phenomena, in which the \emph{strength}, the \emph{range} and even the \emph{sign} of the nonlocal interaction kernel can be widely tuned with high accuracy. To demonstrate this potential, we present numerical results for the propagation of cw laser light and show that paradigm phenomena, such as optical solitons and modulational instabilities (see Fig.\ref{fig1}c) are observable with current experimental capabilities. 

Consider first the propagation of a beam with amplitude $\Omega_{\rm p}$ (see Fig.\ref{fig1}b) and wavenumber $k$ as described by the paraxial wave equation 
\begin{equation}\label{eq1}
 \left(-\frac{i}{2k}\nabla_\perp^2+\frac{\partial}{\partial z}\right)\Omega_{\rm p}({\bf r})=\frac{ik}{2}\chi({\bf r})\Omega_{\rm p}({\bf r}),
\end{equation}
where $\nabla_\perp$ accounts for the transverse dynamics with respect to the axial coordinate ${\bf r}_{\perp}=(x,y)$ perpendicular to the propagation direction $z$.
The relevant medium properties are contained in the complex susceptibility 
\begin{equation}\label{eq2}
\chi=\chi_{\rm R}+i\chi_{\rm I}=\frac{2\wp_{12}^2}{\hbar\epsilon_0\Omega_{\rm p}}\rho_{12},
\end{equation}
which is determined by the dipole matrix element $\wp_{12}$ of the probe transition and the corresponding atomic coherence density $\rho_{12}$. The probe field, $\Omega_{\rm p}$, drives the lower transition between the ground state, $|1\rangle$, and a low-lying excited state, $|2\rangle$, of ladder-type three-level atoms (see Fig.\ref{fig1}a), whose optical response is controlled by a strong control field, driving the upper transition between $|2\rangle$ and a Rydberg state, $|3\rangle$, with a Rabi frequency $\Omega_{\rm c}>\Omega_{\rm p}$. Without interactions, this yields a perfect EIT medium, in which each of the $N$ atoms in the gas settles into a dark state $|d_i\rangle\propto \Omega_{\rm c}|1_i\rangle-\Omega_{\rm p}({\bf r}_i)|3_i\rangle$ ($i=1,...,N$) such that $\rho_{12}=\chi=0$ and the probe beam is unaffected by the atomic medium \cite{fim05}. In the presence of strong Rydberg-Rydberg atom interactions the gas dynamics becomes highly correlated due to the resulting level shifts of multiply excited Rydberg states. Within a critical blockade radius $R_{\rm c}$ all but a single Rydberg excitation are inhibited \cite{lufl01} (see Fig.\ref{fig1}b) and removed from two-photon resonance, thereby diminishing EIT, and, thus giving rise to nonlocal absorption and refraction within a range $\sim R_{\rm c}$. Since the Rydberg state population in the unperturbed dark states $|d_i\rangle$ is proportional to $\Omega_{\rm p}({\bf r}_i)^2$ one, hence, expects an intensity-dependent, i.e. nonlinear, optical response.

Having established a simple picture of the basic mechanisms we now derive the resulting optical susceptibility from the underlying Heisenberg equations for the atomic transition operators $\hat{\sigma}_{\alpha\beta}^{(i)}=|\alpha_i\rangle\langle\beta_i|$ ($\alpha,\beta=1,2,3$). In the limit of low probe intensities ($\Omega_{\rm p}({\bf r}_i)\ll\Omega_{\rm c}$) these can be expanded in $\Omega_{\rm p}/\Omega_{\rm c}$ \cite{fllu02}. Upon adiabatic elimination of $\hat{\sigma}_{12}^{(i)}$ one obtains a single dynamical equation for the two-photon transition operator of the $i$th atom
\begin{equation}\label{eq3}
\frac{d}{dt}\hat{\sigma}_{13}^{(i)}=-\Omega_{\rm c}\frac{\Omega_{\rm p}({\bf r}_i)+\Omega_{\rm c}\hat{\sigma}_{13}^{(i)}}{2\Gamma}-\frac{\gamma_{13}}{2}\hat{\sigma}_{13}^{(i)}-i\sum_{j\neq i}V_{ij}\hat{\sigma}_{33}^{(j)}\hat{\sigma}_{13}^{(i)},
\end{equation}
where $\hat{\sigma}_{33}^{(i)}=\hat{\sigma}_{31}^{(i)}\hat{\sigma}_{13}^{(i)}$, $\Gamma=\gamma+\gamma_{12}-i\Delta$, $\Delta/2$ is the single-photon detuning and the rates $\gamma$, $\gamma_{12}$ and $\gamma_{13}$ account for the spontaneous decay of the intermediate state as well as the linewidth of the probe and two-photon transition, respectively. The last term in eq.(\ref{eq3}) describes the interactions between atoms in the Rydberg state $|3\rangle$ and $V_{ij}=C_6/|{\bf r}_i-{\bf r}_j|^6$ denotes the corresponding van der Waals potential for atoms at positions ${\bf r}_i$ and ${\bf r}_j$. Since the van der Waals coefficient $C_6\propto n^{11}$ drastically increases with the atom's principal quantum number $n$, the interaction between highly excited Rydberg atoms exceeds that of the two low-lying states by many orders of magnitude. Proper inclusion of the resulting strong atomic correlations requires knowledge of the two-body correlators $\hat{\sigma}_{\alpha\beta}^{(j)}\hat{\sigma}_{\alpha^{\prime}\beta^{\prime}}^{(i)}$ whose dynamics follows from eq.(\ref{eq3}) by applying the chain rule. Being primarily interested in the leading order nonlinear contribution to $\chi$, we can once more expand the resulting two-body equations to leading order in $\Omega_{\rm p}$. This amounts to dropping direct three-body correlators and, thus, yields a closed set of evolution equations for the one- and two-body operators.  Setting $\tfrac{d}{dt}\langle\hat{\sigma}_{\alpha\beta}^{(i)}\rangle=\tfrac{d}{dt}\langle\hat{\sigma}_{\alpha\beta}^{(j)}\hat{\sigma}_{\alpha^{\prime}\beta^{\prime}}^{(i)}\rangle=0$, the steady state expectation values are then readily obtained from the resulting set of algebraic equations. Finally, we take the continuum limit by defining continuous densities $\rho_{\alpha\beta}({\bf r})=\sum_i\langle\hat{\sigma}_{\alpha\beta}^{(i)}\rangle\delta({\bf r}-{\bf r}_i)$, and obtain 
\begin{eqnarray}\label{eq4}
\rho_{12}({\bf r})&\!=&\!\frac{i\gamma_{13}\Omega_{\rm p}{(\bf r)}}{\Omega_{\rm c}^2+\gamma_{13}\Gamma}\rho-\frac{\Omega_{\rm p}{(\bf r)}\Omega_{\rm c}^4}{(\Omega_{\rm c}^2-\gamma_{13}\Gamma)|\Omega_{\rm c}^2+\gamma_{13}\Gamma|^2}\rho^2\nonumber\\
&&\times\int d{\bf r'}\:\frac{2|\Omega_{\rm p}({\bf r}^{\prime})|^2V({\bf r}-{\bf r}^{\prime})}{\Omega_{\rm c}^2+\gamma_{13}\Gamma+i\Gamma V({\bf r}-{\bf r}^{\prime})},
\end{eqnarray}
where $\rho=\sum_i\delta({\bf r}-{\bf r}_i)$ is the total atomic density. Together with eq.(\ref{eq2}) this yields the leading-order nonlinear susceptibility and permits to propagate the probe beam according to eq.(\ref{eq1}).

\begin{figure}
\begin{center}
\includegraphics[width=0.95\columnwidth]{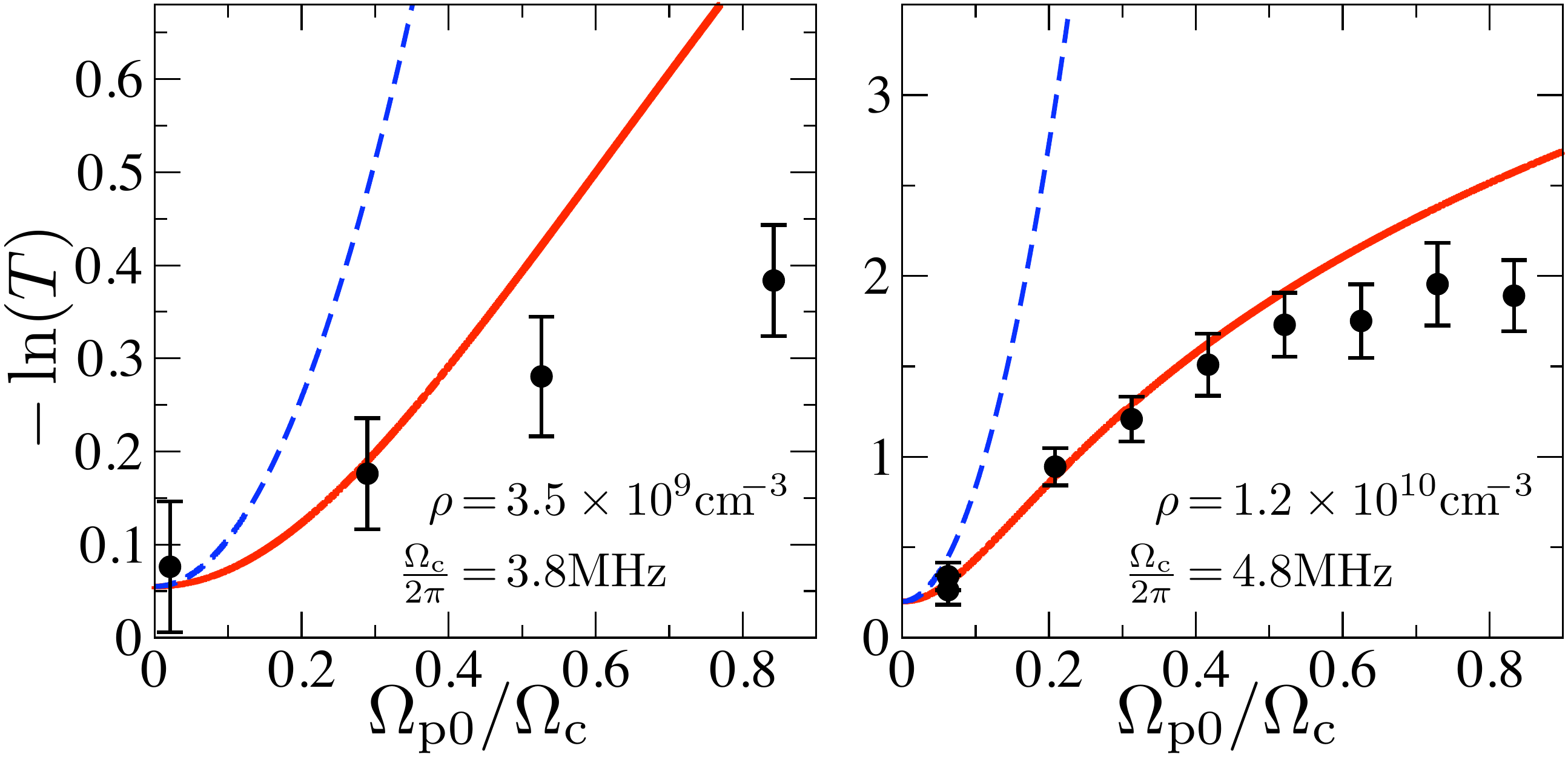}
\caption{Nonlinear transmission of a cold Rubidium Rydberg-EIT medium with $|3\rangle=|60S_{1/2}\rangle$ at two different densities and control Rabi frequencies, and for $\gamma_{12}/2\pi=110$kHz and $\gamma_{13}/2\pi=220$kHz \cite{prma10}.  Up to $\Omega_{{\rm p}0}\approx0.3\Omega_{\rm c}$ there is good agreement between our low-$\Omega_{\rm p}$ prediction eqs.(\ref{eq5}) and (\ref{eq6}) (solid line) and the experimental data \cite{prma10,csa} (symbols). The dashed lines neglect the drop in absorption due to attenuation and averaging over the initial transverse beam profile.}
\label{fig2}
\end{center}
\end{figure}

If the atoms are driven on single-photon resonance, $\Delta=0$, the main interaction effect will be nonlinear absorption. Hence, one can neglect the transverse beam dynamics ($\nabla_{\perp}^2$) as well as the nonlocality in eq.(\ref{eq4}), by setting $\Omega_{\rm p}^2({\bf r}^{\prime})\approx \Omega_{\rm p}^2({\bf r})$. With this simplification one obtains local first and third order susceptibilities, defined by $\chi({\bf r})=\chi^{(1)}+\chi^{(3)}\Omega_{\rm p}^2({\bf r})$. The remaining spatial integral in eq.(\ref{eq4}) can be carried out analytically to give
\begin{eqnarray}\label{eq5}
 \chi^{(1)}_{\rm R}&\!=&\!0\;,\;\; \chi^{(1)}_{\rm I}=\frac{6\pi\gamma\gamma_{13}}{k^3(\gamma_{13}\Gamma+\Omega_{\rm c}^2)}\rho\:,\\
\chi^{(3)}_{\rm R}&\!=&\!-\frac{4\sqrt{2}\pi^3\gamma\Omega_{\rm c}^4C_6|C_6|^{-1/2}}{k^3\sqrt{\Gamma}[\gamma_{13}\Gamma+\Omega_{\rm c}^2]^{7/2}}\rho^2\;,\;\; \chi^{(3)}_{\rm I}=|\chi^{(3)}_{\rm R}|\;.\nonumber
\end{eqnarray}
This expression permits a simple interpretation, by introducing the resonant blockade radius $\tilde{R}_c$, defined by the distance at which the interaction $|C_6|/\tilde{R}_{\rm c}^6$ exceeds the width $\tilde{\delta}_{\rm EIT}=\Omega^2_2/\Gamma$ of the EIT window \cite{gof11}. Substitution of $C_6$ by $\tilde{R}_{\rm c}=(|C_6|/\tilde{\delta}_{\rm EIT})^{1/6}$ shows that $\chi^{(3)}$ is proportional to the corresponding two-level response times the number $R_{\rm c}^3\rho$ of blockaded atoms, which is consistent with the simple picture outlined above and the numerical findings of \cite{atse11}.

Experimentally, nonlinear absorption has been recently studied in a cold Rubidium gas involving $|3\rangle=|60S_{1/2}\rangle$ Rydberg states \cite{prma10}. In the experiments the transmission, $T$, of a Gaussian probe beam ($\Omega_{\rm p}=\Omega_{\rm p0}e^{-r_{\perp}^2/w^2}$) through the gas of length $l$ was measured for different intensities and atomic densities. Within the local approximation this configuration permits  a simple solution of eq.(\ref{eq1}) for the integrated beam transmission
\begin{equation}\label{eq6}
T=T_0\frac{\ln(1+p)}{p}\;,
\end{equation}
where $p=\Omega_{{\rm p}0}^2\chi_{\rm I}^{(3)}(1-T_0)/\chi_{\rm I}^{(1)}$ and $T_0=e^{-k\chi^{(1)}l}$ is the first order transmission. Fig.\ref{fig2} shows a comparison to the measured transmission for two different densities and demonstrates good agreement, even for rather large probe Rabi frequencies of up to $\Omega_{{\rm p}0}\approx0.3\Omega_{\rm c}$. Note that the backaction of the nonlinear beam attenuation onto susceptibility and, equally important, the averaging over the transverse beam profile are both essential for a proper description of the experiment. Neglecting these effects yields the dashed lines in Fig.\ref{fig2}, which significantly overestimates the nonlinear absorption.

\begin{figure}
\begin{center}
\includegraphics[width=0.99\columnwidth]{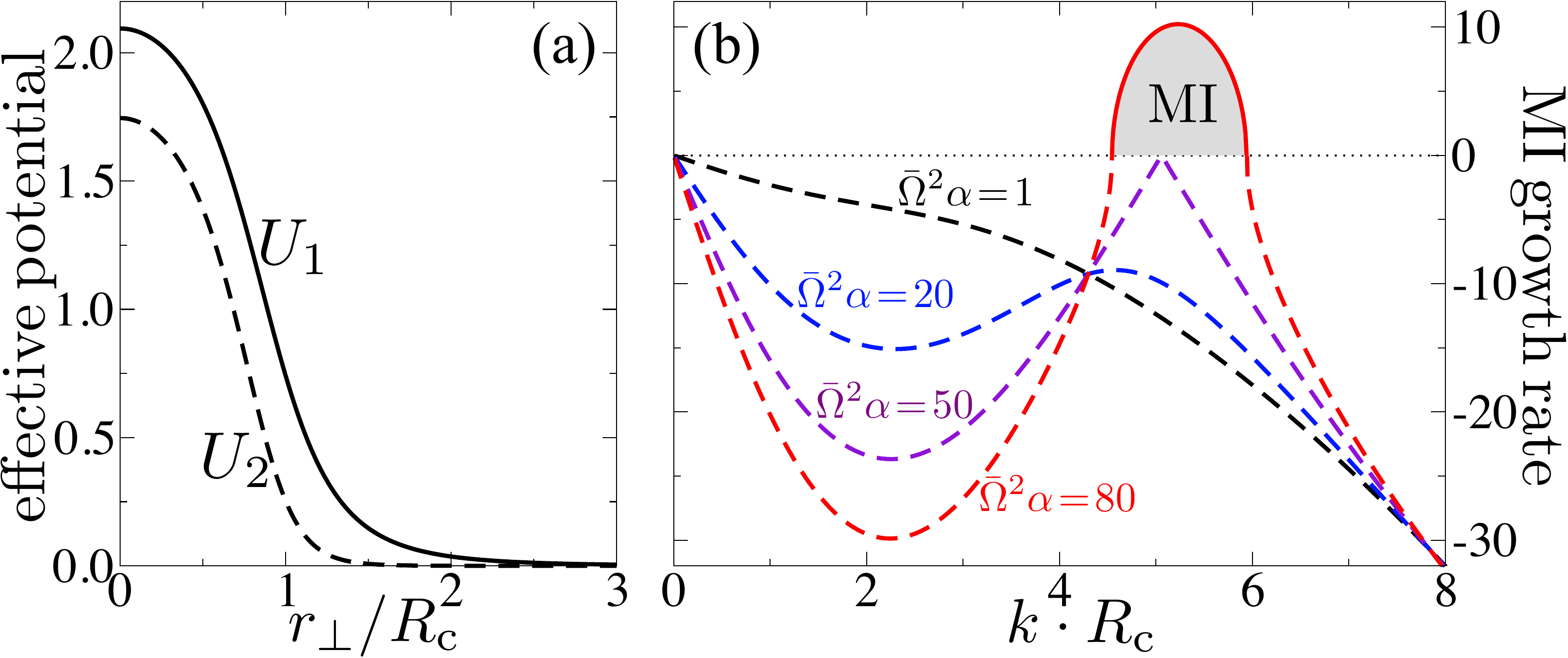}
\caption{(a) Effective photon-photon interaction potentials, introduced in eq.(\ref{eq8}). (b) Growth rate $\Gamma_{\rm MI}$ of intensity modulations with wavenumber $k$ for \emph{defocussing} nonlinearities of different strengths $\Omega^2\alpha$. The dashed lines show the corresponding imaginary part while the solid lines correspond to the real part of $\Gamma_{\rm MI}$. The critical value of $\Omega^2\alpha_{\rm MI}=50.06$ marks the onset of a modulational instability within a narrow window of wave numbers indicated by the grey shaded area for $\Omega^2\alpha=80$.\label{fig3}}
\end{center}
\end{figure}

Since on resonance $\chi^{(3)}_{\rm I}=|\chi^{(3)}_{\rm R}|$ [cf.\ eq.(\ref{eq5})], large nonlinear refraction is inevitably accompanied by high photon loss. However, for large single-photon detunings $\Delta\gg\gamma$ eq.(\ref{eq4}) yields $\chi_{\rm I}\sim(\gamma/\Delta)\chi^{(3)}_{\rm R}$, such that dissipative loss can be greatly suppressed. For instance, for a Rubidium Rydberg gas with $\Omega_{\rm c}/2\pi=5$MHz, $\Omega_{\rm p}/2\pi=0.5$MHz, $\Delta=30$GHz and $\rho=8\times10^{13}$cm$^{-3}$ one obtains a large absorption length of $l_{\rm abs}\approx1$mm, and yet a high nonlinear refractive index $n_{2}\approx2\times10^4$cm$^2$/W which is 5 orders of magnitude greater than previously obtained with ultracold Rb groundstate atoms at the same density \cite{hhd99}.

\begin{figure}[b]
\includegraphics[width=0.99\columnwidth]{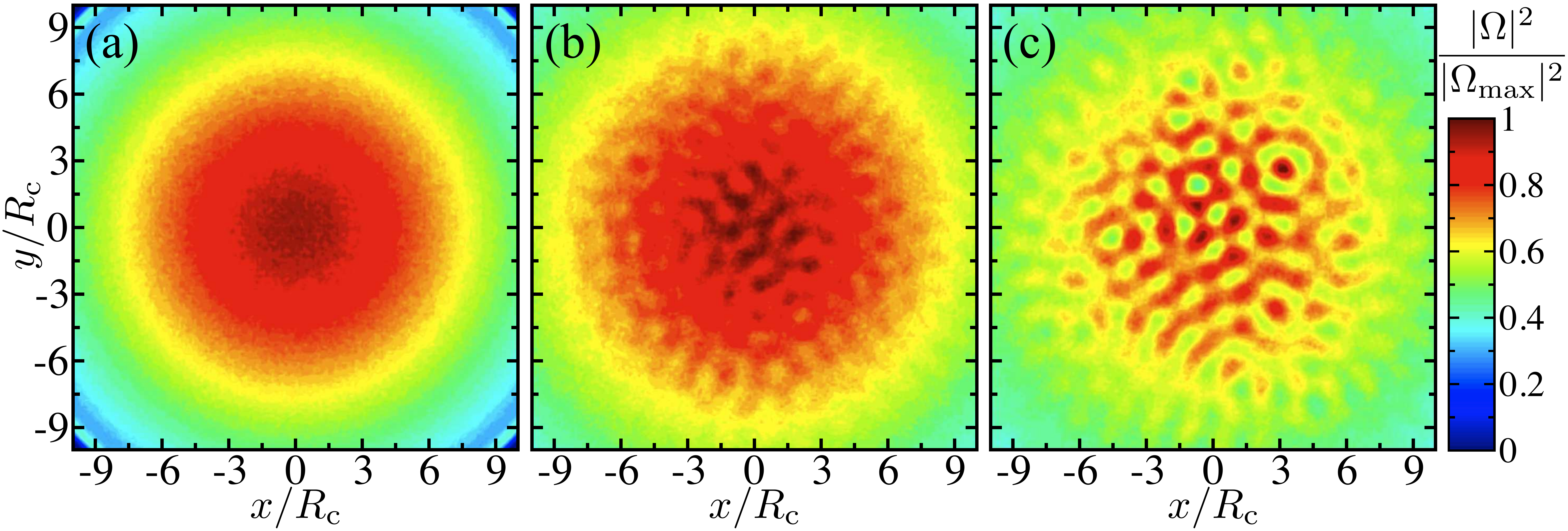}
\caption{Output beam profile $|\Omega|^2(\v r_\perp)$ for $\Omega_{\rm p0}/2\pi=0.35$MHz, $\Omega_{\rm c}/2\pi=80$MHz and $\Delta/2\pi=1.2$GHz after propagation over $l=210\mu$m through a Rb($70S_{1/2}$) EIT medium at three different densities (a) $4\times10^{13}$cm$^{-3}$, (b) $5.5\times10^{13}$cm$^{-3}$ and (c) $8\times10^{13}$cm$^{-3}$. For the color coding each distribution has been normalized by the actual maximum intensity $|\Omega_{\rm max}|^2$.}
\label{fig4}
\end{figure}

As refraction starts to dominate absorption, the nonlocality of $\chi^{(3)}$ [cf. eq.(\ref{eq4})] becomes significant. To account for its effects on the transverse beam propagation we recast eqs.(\ref{eq2},\ref{eq4}) into
\begin{equation}
\chi({\bf r})=-\frac{12\pi\gamma\rho^2}{k^3\Delta\Omega_{\rm c}^2}\int d{\bf r'}\frac{|\Omega_{\rm p}({\bf r}^{\prime}_{\perp},z)|^2}{1+\tfrac{|r^{\prime}-r|^6}{R_{\rm c}^6}}-i\frac{\gamma}{\Delta}\frac{|\Omega_{\rm p}({\bf r}^{\prime}_{\perp},z)|^2}{\left[1+\tfrac{|r^{\prime}-r|^6}{R_{\rm c}^6}\right]^2}\;,
\end{equation}
where we assumed $\gamma_{12}\ll\gamma$, $\gamma_{13}\ll\delta_{\rm EIT}=\Omega_{\rm c}^2/\Delta$ and introduced the off-resonant blockade radius $R_{\rm c}=(C_6/\delta_{\rm EIT})^{1/6}$ ($C_6\Delta>0$) \cite{gof11} set by the off-resonant EIT width $\delta_{\rm EIT}$. 
To simplify matters, we proceed by defining scaled coordinates $\tau=z/(kR_{\rm c}^2)$, ${\bm{\xi}}={\bf r}_{\perp}/R_{\rm c}$ and the dimensionless probe amplitude $\Omega$, normalized to $\int \Omega^2({\bm \xi },\tau)d^2\xi=1$. Retaining the local approximation along the propagation direction \footnote{We verified numerically that the local approximation still holds along the propagation direction for the parameters of this work.} this yields a two-dimensional nonlinear Schr\"odinger equation
\begin{eqnarray}\label{eq8}
 i\partial_\tau{\Omega}({\bm\xi},\tau)\!&\!=\!&\!\left[-\frac{\nabla_{\bm\xi}^2}{2}+\alpha\!\int d{\bm\xi}^{\prime}|\Omega({\bm\xi^{\prime}},\tau)|^2U_1({\bm\xi}-{\bm\xi}^{\prime})\right.\\
 &&\left. - i\frac{\gamma}{\Delta}\alpha\!\int d{\bm\xi}^{\prime}|\Omega({\bm\xi^{\prime}},\tau)|^2U_2({\bm\xi}-{\bm\xi}^{\prime})\right]\Omega({\bm\xi},\tau),\nonumber
\end{eqnarray}
where $\alpha=\tfrac{36\pi^2\rho^2\gamma^2P_{\rm p}}{\hbar k^4cR_{\rm c}^3\Omega_{\rm c}^4}C_6$ parametrizes the strength of the nonlinearity, $P_{\rm p}$ denotes the probe beam power and the effective interaction potentials $U_m(\xi) = \int_{-\infty}^{\infty}\d z\left[1+(\xi^2+z^2)^3\right]^{-m}$ are shown in Fig.\ref{fig3}a. As $\alpha\propto C_6$, repulsive atomic interactions lead to self-defocussing nonlinearities, while attractive atomic interactions map onto self-focussing nonlinearities.

The former case, can, e.g., be realized with cold Rb$(nS_{1/2})$ Rydberg states as in the experiments  \cite{prma10} discussed above. While the resulting photon-photon interactions are isotropically repulsive, the corresponding momentum space interaction $\tilde{U}_1(k)$ is not sign-definite. 
Despite being \emph{defocussing}, the present nonlocal interaction can, consequently, promote a modulational instability. In Fig.\ref{fig3}b we show the corresponding rate $\Gamma_{\rm MI}(k) = -\tfrac{k}{2}\sqrt{k^2 -4\alpha\Omega^2 \tilde{U}_1(k)}$ \cite{wkb02} for a given mode with wave number $k$ to grow out of a homogenous amplitude $\Omega({\bm\xi})={\rm const.}$. This growth rate $\Gamma_{\rm MI}$ assumes real values within a narrow range around $k\approx 2\pi$ at a critical interaction strength $\alpha\Omega^2\approx50$ \cite{hnp10}, resulting in stable transverse intensity modulations on a length scale $\sim R_{\rm c}$ as the beam propagates through the medium (cf. Fig.\ref{fig1}c). To examine their observability, we performed numerical simulations of eq.(\ref{eq8}) for a Rb($70S_{1/2}$) Rydberg gas traversed by a super-Gaussian beam $\Omega=\Omega_0 e^{-(\xi/w)^\nu+i\phi}$ with $\nu=6$ and small spatial phase noise $\phi$. Fig.\ref{fig4} shows calculated output intensity profiles for different atomic densities and demonstrates that highly localized, rather regular intensity patterns can be realized in high density gases with feasible laser parameters. 

\begin{figure}[t]
\includegraphics[width=.99\columnwidth]{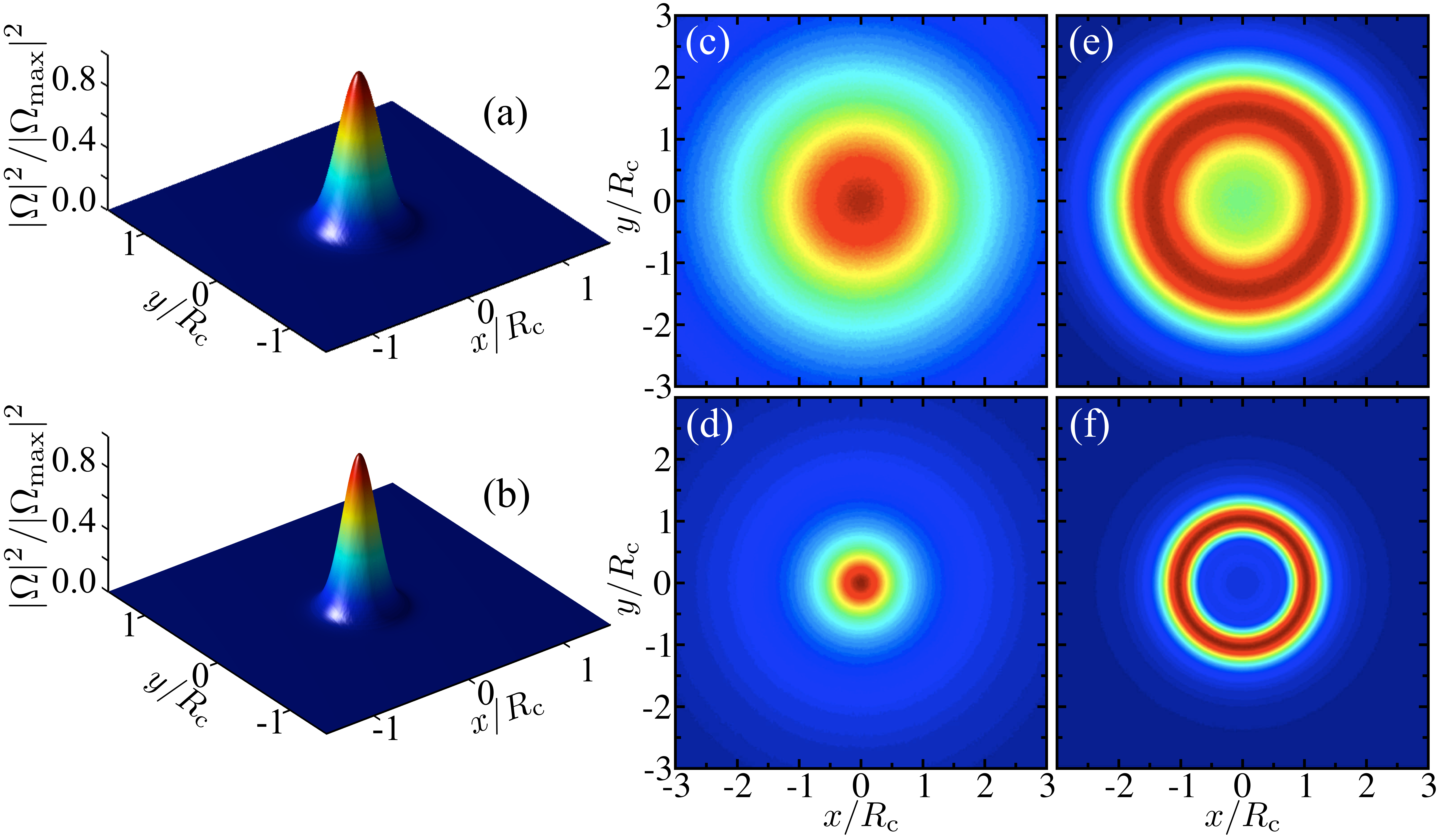}
\caption{Intensity profiles $|\Omega|^2(\v r_\perp)$ for a Sr($50^1S_0$) EIT medium with $\Omega_0/2\pi=0.3$MHz, $\Omega_{\rm c}/2\pi=15$MHz, $\Delta/2\pi=3.2$GHz at two different densities of (a,c,e) $8\times10^{11}$cm$^{-3}$ and  (b,d,f) $1.2\times10^{12}$cm$^{-3}$. Panels (a,b) show the stable soliton solutions, while panels (c-f) show the compressed output intensity profile after a propagation length $l=240\mu$m for an input beam with (c,d) $\nu=2$ and (e,f) $\nu=6$. The color coding is identical to Fig.\ref{fig4}.}
\label{fig5}
\end{figure}

Self-focussing nonlinearities arise from attractive Rydberg interactions, as occurring between $n^1S_0$ states of Strontium atoms \cite{mmn11}, for which EIT has been recently observed \cite{mami07}. In this case, modulational instabilities can, in principle, occur for any $\alpha<0$ and sufficiently large beam widths. More importantly, however, the attractive photonic softcore-interaction enables the formation of \emph{stable} bright solitons, leading to tight beam focussing. A simple variational analysis of eq.(\ref{eq8}) (see, e.g., \cite{mhk11}) yields a critical interaction strength of $\alpha_{\rm so}\approx0.71$, above which stable bright solitons exist. Figs.\ref{fig5}a and \ref{fig5}b show two examples for a Sr($50^1S_0$) gas and reveal a characteristic soliton size $\lesssim R_{\rm c}$. In addition we show final intensity profiles for a Gaussian ($\nu=2$) and super-Gaussian ($\nu=6$) input beam. All cases are for an input width of $w=3R_c$ and demonstrate significant focussing after the considered propagation length $l=240\mu$m. In the latter case, the beam compression is superimposed by a radial MI leading to a ring-shaped hollow output beam. With a typical size of several $\mu$m these structures are readily observable experimentally.

In conclusion, we have presented a theory for the nonlinear response of a strongly interacting Rydberg-EIT gas, giving good agreement with recent  measurements. The derived expressions for the third-order susceptibility suggest that huge nonlinearities of highly nonlocal character can be experimentally realized, which provides an ideal setting to study complex nonlinear wave phenomena. To demonstrate these prospects we have shown that the observation of basic effects such as the formation of bright solitons and collapse-arrested modulational instabilities are within experimental reach. The latter is particularly interesting in the uncommon case of defocussing, nonlocal nonlinearities. Here, the repulsion between emerging intensity peaks combined with transverse beam confinement may promote the formation of transverse supersolid or crystalline states of photons. This question may be addressed within the present approach, extended to quantum light in order to account for atom-photon and photon-photon correlations, which would open up a general framework for studying many-body physics with strongly interacting photons. From a different perspective, we expect the discussed nonlinear light propagation to be relevant for interpreting cold Rydberg gas experiments at high densities.

We are grateful to A.~Gorshkov, M.D.~Lukin, J.D.~Pritchard, C.S.~Adams, S.~Skupin, J.~Otterbach and M.~Fleischhauer for valuable discussions, and thank C.S.~Adams for providing unpublished experimental data.

\end{document}